\documentclass[]{article}

\usepackage{amssymb}
\usepackage{graphicx}\usepackage{amsmath,amssymb}
\usepackage{color}
\usepackage[english]{babel}
\usepackage{xcolor}
\usepackage{srcltx}
\usepackage{greekbf}
\usepackage{nicefrac}
\usepackage{empheq}
\usepackage{enumerate}
\usepackage{wrapfig}
\usepackage{amsthm}

\usepackage{multicol}

\usepackage{mathtools}
\mathtoolsset{showonlyrefs}
    \mathtoolsset{centercolon}
\usepackage{mathrsfs}
\usepackage{epstopdf}

\allowdisplaybreaks

\DeclareMathAlphabet{\mathbf}{OT1}{cmr}{bx}{it}
\DeclareMathAlphabet{\mathssb}{OT1}{cmss}{bx}{n}
\DeclareMathAlphabet{\mathssn}{OT1}{cmss}{m}{n}
\DeclareMathAlphabet{\mathub}{OT1}{cmr}{b}{n}

\DeclareMathAlphabet{\mathpzc}{OT1}{pzc}%
                                 {m}{it}

\theoremstyle{definition}

\newcommand{\bydef}{\,\raise.050ex\hbox{\rm:}\kern-.025em\hbox{\rm=}\,}
\newcommand{\defby}{=\raise.075ex\hbox{\kern-.325em\hbox{\rm:}}\,}

\def\qed{\relax\ifmmode\hskip2em \Box\else\unskip\nobreak\hskip1em $\Box$\fi}
\newcommand {\Ec}  {\mathcal{E}}
\newcommand {\Fc}  {\mathcal{F}}

\newcommand {\Kc}  {\mathcal{K}}

\newcommand {\Lc}  {\mathcal{L}}

\newcommand {\Nc}  {\mathcal{N}}

\newcommand {\Sc}  {\mathcal{S}}

\newcommand {\Uc}  {\mathcal{U}}

\newcommand {\Wc}  {\mathcal{W}}

\newcommand {\ab} {\mathbf{a}}
\newcommand {\bb} {\mathbf{b}}
\newcommand {\cb} {\mathbf{c}}

\newcommand {\eb} {\mathbf{e}}
\newcommand {\fb} {\mathbf{f}}
\newcommand {\gb} {\mathbf{g}}

\newcommand {\mb} {\mathbf{m}}
\newcommand {\nb} {\mathbf{n}}
\newcommand {\pb} {\mathbf{p}}
\newcommand {\qb} {\mathbf{q}}

\newcommand {\ub} {\mathbf{u}}
\newcommand {\vb} {\mathbf{v}}
\newcommand {\wb} {\mathbf{w}}

\newcommand {\Ab} {\mathbf{A}}

\newcommand {\Cb} {\mathbf{C}}

\newcommand {\Gb} {\mathbf{G}}

\newcommand {\Kb} {\mathbf{K}}

\newcommand {\Ib} {\mathbf{I}}

\newcommand {\Mb} {\mathbf{M}}

\newcommand {\Pb} {\mathbf{P}}
\newcommand {\Qb} {\mathbf{Q}}

\newcommand {\Ro} {\mathbb{R}}

\newcommand {\etab}      {\mathbf{\eta}}

\newcommand {\chib}      {\mathbf{\chi}}

\newcommand{\sym}{\mathop{\mathrm{sym}}}

\newcommand{\ov}{\overline}

\begin{document}

\begin{center}
 {\bf \Large
A Nonlinear Theory of\\ Prestressed Elastic Stick-and-Spring Structures}
\end{center}
\medskip

\begin{center}
 {\large
Antonino Favata$^\natural$, Andrea Micheletti$^\sharp$, 
Paolo Podio-Guidugli$^\sharp$}
\end{center}

\begin{center}
 {
$^\natural$Institute of Continuum Mechanics and Material Mechanics,\\ Hamburg University of Technology\footnote{Ei\ss endorfer Stra\ss e 42, 21073 Hamburg Germany. \\antonino.favata@tuhh.de.}
}
\end{center}

\begin{center}
 {
$^\sharp$ Department of Civil Engineering and  Information Engineering\\
University of Rome To{\color{black}rV}ergata\footnote{Via Politecnico 1, 00133, Rome, Italy. \\micheletti@ing.uniroma2.it, ppg@uniroma2.it}
}
\end{center}

\begin{abstract}
The discrete modeling of a large class of mechanical structures can be based on a stick-and-spring concept. We here present a stick-and-spring  theory with potential application to the statics and the dynamics of such nanostructures as graphene, carbon nanotubes, viral capsids, and others. A key feature of our theory is its geometrical nonlinearity: we combine  exactly defined strain measures with a general linear stress response; another, rarely found feature is a careful account of prestress states. A linear version is firstly proposed, where attention is restricted to study small displacements from an unstressed reference placement. Next, a theory linearized about a prestressed (preloaded or not) placement is displayed, which is based on a careful analysis of the tangent stiffness operator and its two parts, the elastic and prestress stiffness operators. Finally, two examples are proposed and solved; when an analytical solution is of prohibitive complication, numerical solutions are given, by the use of a specifically implemented `stick-and-spring' code.
\end{abstract}


\begin{flushright}
\textit{Dedicated to Professor Leonid M. Zubov on the occasion of his 70th birthday.}
\end{flushright}

\section{Introduction}
In this paper we propose a theory of elastic molecular structures that may be in a stressed state before their static response to applied loads and imposed nodal displacements is determined, or their free vibrations are studied. In addition to hoping it to be of interest \textit{per se}, we believe that this theory  is applicable in many contexts, both statical and dynamical, where the objects whose mechanical behavior is under scrutiny are carbon allotropes (nanotubes, graphene sheets, nanoribbons), molecules, protein complexes,  {\em{et cetera}}.

 We regard a molecular structure as a collection of {\emph{nodes}}, {\emph{edges}}, and {\emph{wedges}}: edges are imagined as node-to-node inflexible but extensible straight {\em sticks}, acting as {\em axial springs} when extended; wedges are imagined as complexes of two sticks sharing one end node, equipped by a {\em torsion spring} reacting to relative rotations of the wedge sticks in their common plane. Accordingly, we  call the structures we model {\em Stick-and-Spring structures} (S\&S).
 Similar but simpler molecular structures are the so-called \emph{elastic networks}, that is, roughly speaking, sets of nodes pair-wise connected by edges,
 a concept used to model generally small displacements from an equilibrium configuration of protein complexes (see e.g. \cite{Ba}). Another class of molecular structures comprises the {\em molecular dynamics models}, where a two- or three-body potential accounts for the interactions among particles whose relative distance is less than a prescribed cut-off length \cite{Ab,Br,Br2,Fin,Ter}.
 %

 Stick-and-spring models have been widely used to predict the mechanical behavior of Carbon NanoTubes (CNTs) and graphene. The linear model exploited in \cite{CG} to obtain closed-form expressions for the elastic properties of armchair and zigzag CNTs has been extended in \cite{Xiao} to study torsion loading and handle  nonlinearities, by means of a modified Morse potential.  A similar approach is used in \cite{SL} to investigate various loading conditions, and in \cite{Wang} to evaluate  effective in-plane stiffness and bending rigidity of armchair and zigzag CNTs. Molecular mechanics has also been employed as a scale-bridging method to build shell theories \cite{BFPG,Cha1,FPG}. In \cite{CGG1}, the model of \cite{CG} is extended to chiral CNTs, an issue addressed also in \cite{CGG}. Recently,  a new formulation of the stick-and-spring model has been proposed \cite{Merli}, allowing for general load conditions, arbitrary chirality, and an initial stressed state.   It is worth mentioning the computational methods presented in \cite{Meo}, where non-linear torsional spring elements are adopted and implemented in a Finite Element (FE) code. The mechanical properties of graphene sheets and nanoribbons have been analyzed with a similar approach; in particular, FE formulations, employing both linear \cite{Gerg1} and non-linear springs \cite{GC,Gerg,Gian,Gian1}, have been given.

The method of structural analysis we propose is fairly general and versatile. Although it can be profitably adopted for all of the above mentioned applications (and many others),  here we only give a description of its general features, illustrated by an analysis of simple concept structures of no special applicative significance, except for the cyclohexane isomers considered in Remark 4, Section \ref{presss}; real applications will be dealt with elsewhere. The main novelty of our approach consists in handling prestress without neglecting geometrical nonlinearities; a careful account of prestress is important in various contexts where our theory is potentially applicable: see e.g.  \cite{SWG} for globular proteins, and \cite{SPPG}, where certain aspects of the mechanical phenomenolgy of graphene and CNTs  are investigated.

Here is a summary of the contents of our paper. In Section 2, we describe the topology and the kinematics of stick-and-spring structures; in particular, we define \textit{exact} strain measures and derive their \textit{linearized} version. 
Then, in Section 3, we use a virtual power argument to derive the nodal equations that must be satisfied by all stress states compatible with the data, that is, balancing the applied load in the given reference placement. These equations implicitly define the {\em equilibrium operator} and its transpose, the {\em kinematic compatibility operator}. Just as for any other discrete structural system, the dimensions of the null spaces of these operators provide a useful typological criterion for the structural systems under study: e.g., the set of admissible self-stress states can be identified with the nullspace of the equilibrium operator. A linear theory of elastic S\&S structures consists in coupling the nodal equilibrium equations with a linear constitutive equation delivering edge and wedge stresses in terms of linearized edge and wedge strain measures; such an equation is tantamount of specifying the 
{\em linear stiffness operator}. 

However, the effects of prestress on a system's response cannot be predicted within a fully linear setup.  To account for these effects,  in Section 4 we derive an expression for the {\em tangent stiffness operator}, that is, the Hessian of the stored energy. This operator admits an additive decomposition into an {\em elastic stiffness operator} (the analog of the stiffness operator encountered in the linear theory) plus a {\em prestress stiffness operator}.  While the former operator accounts for first-order changes in the strain measures, the latter accounts for orientational changes of edges and wedges and for the stresses they carry; thus, its contribution to balancing the service loads can be nonnull even when the accompanying displacements entail null linearized strains. Such contribution is especially important when it comes to assessing the stability of an equilibrium placement, which is connected with the sign-definiteness of the tangent stiffness operator, when seen as a quadratic form. In this connection, we borrow from rigidity theory \cite{C99} (see also \cite{M13}) the notions of {\em prestress stability} and {\em super-stability}, and we restate them for S\&S structures.

Knowledge of the equilibrium and tangent stiffness operators allows for application of standard numerical machineries to solve a number of structural problems, such as computing load-displacement paths, performing buckling analyses, determining natural frequencies and vibration modes, and integrating the nonlinear motion equations. In Section 5, two simple concept structures are analyzed, in the linear, linearized, and nonlinear setups of our theory; in particular,  the numerical results obtained by the use of a specifically implemented `stick-and-spring' code. Some directions of future research are mentioned in our final Section 6.

\section{Topology and Kinematics}

%
\subsection{Topology}
A S\&S structure is a triplet $\Sc=(\Nc,\Ec,\Wc)$, consisting of: (i) a collection $\Nc$ of $N$ points, called {\em nodes}, of the three-dimensional Euclidean space; (ii) a collection $\Ec$ of $E$ {\em edges}, that is, two-elements subsets of $\Nc$; (iii) a collection $\Wc$ of $W$ {\em wedges}, that is, three-elements subsets of $\Nc$. We say that $ij\in\Ec$ is the edge connecting nodes $i, j\in\Nc$, and that $ijk\in\Wc$, with $i, j, k\in\Nc$, is the wedge with {\em head node} $i$  and {\em tail nodes} $j$ and $k$.

\begin{figure}[h]
\centering
\includegraphics[scale=0.9]{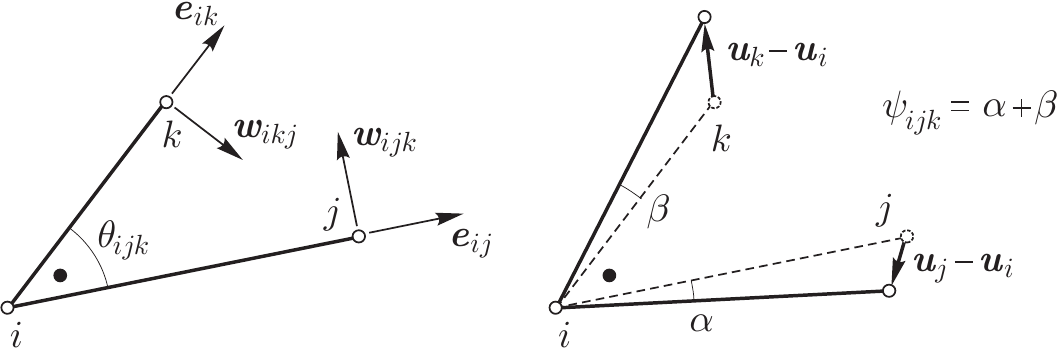}
\caption{A wedge and its change in angle. Here and henceforth nodes are denoted by a circle $\circ$, wedges by a bullet $\bullet$.}
\label{wedge} 
\end{figure}

We choose a referential placement for $\cal S$ in the three-dimensional ambient space, and we denote by $\pb_i$  the \emph{referential position vector} of the typical node $i$ with respect to a chosen origin point. We write $l_{ij}:=|\pb_i-\pb_j|$ for the referential length of edge $ij$, and
\begin{equation}\label{eb}
\eb_{ij}:=\frac{1}{l_{ij}}(\pb_j-\pb_i)=-\eb_{ji}
\end{equation}
for the unit vector directed from node $i$ to node $j$.
%
Moreover, given a wedge $ijk$, we associate with it two unit vectors, namely,
\begin{equation}\label{proj}
\wb_{ijk}=\frac{\Pb_{ij}\eb_{ik}}{|\Pb_{ij}\eb_{ik}|} \quad \mathrm{and} \quad \wb_{ikj}=\frac{\Pb_{ik}\eb_{ij}}{|\Pb_{ik}\eb_{ij}|},
\end{equation}
where $\Pb_{ij}:=\Ib-\eb_{ij}\otimes\eb_{ij}$ and $\Pb_{ik}:=\Ib-\eb_{ik}\otimes\eb_{ik}$ are the orthogonal projectors on the planes of normal $\eb_{ij}$ and $\eb_{ik}$, respectively; note that both $\wb_{ijk}$ and $\wb_{ikj}$ point `inward' (see  Fig.\,\ref{wedge}\,(left)), in the sense that both $\wb_{ijk}\cdot\eb_{ik}>0$ and  $\wb_{ikj}\cdot\eb_{ij}>0$;
note also that, here and henceforth, {\em no summation over repeated indices should be presumed, unless explicitly indicated}. On letting $\vartheta_{ijk}:=\arccos(\eb_{ij}\cdot\eb_{ik})$ be the referential angle between edges $ij$ and $ik$, we have that
\begin{equation}
\sin\vartheta_{ijk}=\Big(1-(\eb_{ij}\cdot\eb_{ik})^2  \Big)^{1/2}=|\Pb_{ij}\eb_{ik}|=|\Pb_{ik}\eb_{ij}|,
\end{equation}
so that equation \eqref{proj} can be rewritten as:
\begin{equation}\label{wbb}
\wb_{ijk}=\frac{\Pb_{ij}\eb_{ik}}{\sin\vartheta_{ijk}} \ \ \mathrm{and} \ \ \wb_{ikj}=\frac{\Pb_{ik}\eb_{ij}}{\sin\vartheta_{ijk}},
\end{equation}
an expression which will be useful later
(for collinear edges, the role of the unit vectors in \eqref{proj} is played by the corresponding limits for $\vartheta_{ijk}\rightarrow 0, \pi$).

%

\subsection{Kinematics, exact and linearized}\label{kinematics}
A S\&S elastic structure stores energy because edges change length and wedges change angle; we then define two associated {\em strain measures}.\footnote{To simulate protein docking, it might be important to account also for changes in the dihedral angle formed by two edge-sharing wedges. For the moment being, we postpone consideration of this mechanism of elastic energy storage.}

Given a set of nodal displacements $\ub_i\;(i\in{\mathcal N})$ for $\mathcal S$, we denote the \textit{current position vector} of node $i$ as
\begin{equation}\label{qb}
\qb_i:=\pb_i+\ub_i;
\end{equation}
the unit vector currently directed from node $i$ to node $j$ is:
\begin{equation}\label{qbb}
\cb_{ij}:=\frac{1}{\ell_{ij}}(\qb_j-\qb_i)=-\cb_{ji}, \quad \ell_{ij}:=|\qb_j-\qb_i|,
\end{equation}
with $\ell_{ij}$ the current length of edge $ij$. Next, given a wedge $ijk$ in its current placement, whose current angle is:
\begin{equation}\label{wa}
\theta_{ijk}:=\arccos(\cb_{ij}\cdot\cb_{ik}),
\end{equation}
introduce the unit vectors
\begin{equation}
\widetilde{\wb}_{ijk}=\frac{\widetilde{\Pb}_{ij}\cb_{ik}}{|\widetilde{\Pb}_{ij}\cb_{ik}|} \quad \mathrm{and} \quad \widetilde\wb_{ikj}=\frac{\widetilde\Pb_{ik}\cb_{ij}}{|\widetilde\Pb_{ik}\cb_{ij}|},
\end{equation}
where $\widetilde\Pb_{ij}:=\Ib-\cb_{ij}\otimes\cb_{ij}$ and $\widetilde\Pb_{ik}:=\Ib-\cb_{ik}\otimes\cb_{ik}$ are the orthogonal projectors on the planes of normal $\cb_{ij}$ and $\cb_{ik}$, respectively.

The \textit{change in length} of edge $ij$ is:
\begin{equation}\label{chle}
\delta l_{ij}:=\ell_{ij} - l_{ij};
\end{equation}
the \textit{change in angle} of wedge $ijk$ is:
\begin{equation}\label{chan}
\delta\theta_{ijk}:=\arccos\,(\cb_{ij}\cdot\cb_{ik})-\arccos\,(\eb_{ij}\cdot\eb_{ik});\footnote{For us,  wedge angles are not greater than $\pi$, both in the reference and in the current placement. An alternative version of \eqref{chan} is arrived at by the use of \eqref{eb} and \eqref{qb}:
\begin{equation}
\delta\theta_{ijk}= \arccos\Big(\frac{l_{ij}\eb_{ij}+\ub_i-\ub_j}{|l_{ij}\eb_{ij}+\ub_i-\ub_j|}\cdot \frac{l_{ik}\eb_{ik}+\ub_i-\ub_k}{|l_{ik}\eb_{ik}+\ub_i-\ub_k|}\Big)-\arccos\,(\eb_{ij}\cdot\eb_{ik}).
\end{equation}
}
\end{equation}
$\delta l_{ij}$ and $\delta\theta_{ijk}$ are the exact strain measures we use.
In a \textit{motion}:
$$
\{t\mapsto\qb_i(t)\,|\, i\in\Nc, t\in T\subset\Ro\},
$$
the {\em realized velocity} at time $t$ of the typical node of $\mathcal S$ is:
$$
\vb_i(t):=\dot{\qb}_i(t).
$$
Easy computations show that the time rates of the strain measures we just introduced are, respectively,
\begin{equation}\label{rcl}
(\delta l_{ij})^{\cdot}=\cb_{ij}\cdot(\vb_i-\vb_j)
\end{equation}
and
\begin{equation}\label{rca}
(\delta\theta_{ijk})^\cdot=-\frac{1}{\ell_{ik}}\,\widetilde\wb_{ikj}\cdot(\vb_k-\vb_i)-\frac{1}{\ell_{ij}}\,\widetilde{\wb}_{ijk}\cdot(\vb_j-\vb_i)\,.
\end{equation}

To deduce the \textit{linear} versions of the exact deformation measures, we take $\displaystyle{\xi:=\sup_{ij\in{\mathcal E}}\, |\ub_i|/l_{ij}}$ as smallness bookkeeping parameter and we introduce the mapping:
\begin{equation}\label{sma}
\xi\,\mapsto \, \qb_i(\xi)=\pb_i+\xi\ub_i\,.
\end{equation}
Firstly, composition of  \eqref{sma} with \eqref{chle} yields:
\begin{equation}
\xi\mapsto\widetilde{\delta l}_{ij}(\xi):=|\pb_j-\pb_i+\xi(\ub_j-\ub_i)|-l_{ij}=\big(\widetilde{\delta l}_{ij}(0)-l_{ij}\big)+\widetilde{\delta l}_{ij}^\prime(0)\,\xi+O(\xi^2);
\end{equation}
as
\[
\widetilde{\delta l}_{ij}(0)=l_{ij}\quad\textrm{and}\quad \widetilde{\delta l}_{ij}^\prime(0)=\eb_{ij}\cdot(\ub_j-\ub_i)=\eb_{ji}\cdot(\ub_i-\ub_j),
\]
we have that 
\begin{equation}\label{linchle}
\widetilde{\delta l}_{ij}(\xi)=\big(\eb_{ji}\cdot(\ub_i-\ub_j)\big)\xi+O(\xi^2);
\end{equation}
hence, we take
\begin{equation}\label{eps}
\varepsilon_{ij}:=\eb_{ji}\cdot(\ub_i-\ub_j)\,,
\end{equation}
as the {\em linear strain measure accounting for length changes}; its time rate is:
\begin{equation}\label{reps}
\dot\varepsilon_{ij}:=\eb_{ji}\cdot(\vb_i-\vb_j)\,.
\end{equation}

 Secondly, this time via composition of \eqref{sma} with \eqref{chan}, we find that the {\em linear strain measure accounting for angle changes} is:
\begin{equation}\label{psi}
\psi_{ijk}:=-\frac{1}{l_{ik}}\,\wb_{ikj}\cdot(\ub_k-\ub_i)-\frac{1}{l_{ij}}\,\wb_{ijk}\cdot(\ub_j-\ub_i)\,,
\end{equation}
or rather, after a rearrangement,
\begin{equation}\label{psi2}
\psi_{ijk}=\left(\frac{\wb_{ijk}}{l_{ij}}+\frac{\wb_{ikj}}{l_{ik}}\right)\cdot\ub_i- \frac{\wb_{ijk}}{l_{ij}}\cdot\ub_j-\frac{\wb_{ikj}}{l_{ik}}\cdot\ub_k\,;
\end{equation}
the relative time rate is:
\begin{equation}\label{rpsi}
\dot\psi_{ijk}=\left(\frac{\wb_{ijk}}{l_{ij}}+\frac{\wb_{ikj}}{l_{ik}}\right)\cdot\vb_i- \frac{\wb_{ijk}}{l_{ij}}\cdot\vb_j-\frac{\wb_{ikj}}{l_{ik}}\cdot\vb_k\,.
\end{equation}

The geometrical meaning of \eqref{psi} becomes clear after a glance to Fig. \ref{wedge} (right): $\psi_{ijk}$ consists of two contributions, $\alpha$ and $\beta$; each of them is found by projecting the relative displacement -- that is, $(\ub_j-\ub_i)$ for $\alpha$ and $(\ub_k-\ub_i)$ for $\beta$ --  in the direction perpendicular to, respectively, $\eb_{ij}$ or $\eb_{ik}$, and by dividing the result by $l_{ij}$  or $l_{ik}$.\footnote{
To prove that \eqref{psi} is the linear version of \eqref{chan}, we compute the linear approximation about $\xi=0$ of the following mapping:
\begin{equation}
\widetilde{\delta\theta}_{ijk}(\xi):=\arccos\left(\frac{\mb(\xi)\cdot\nb(\xi)}{|\mb(\xi)||\nb(\xi)|}\right), \qquad \mb(\xi):=l_{ij}\eb_{ij}+\xi(\ub_j-\ub_i), \quad \nb(\xi):=l_{ik}\eb_{ik}+\xi(\ub_k-\ub_i).
\end{equation}
As
\begin{equation}
\begin{aligned}
-\sin\big(\widetilde{\delta\theta}_{ijk}(\xi)\big)\,\widetilde{\delta\theta}_{ijk}^\prime(\xi)=\Big(\frac{\mb(\xi)\cdot\nb(\xi)}{|\mb(\xi)||\nb(\xi)|}\Big)^\prime=
\frac{\mb'(\xi)\cdot\nb(\xi)+\mb(\xi)\cdot\nb'(\xi)}{|\mb(\xi)||\nb(\xi)|}-\frac{\mb(\xi)\cdot\nb(\xi)}{\mb(\xi)||\nb(\xi)|}\left(\frac{\mb(\xi)\cdot\mb'(\xi)}{|\mb(\xi)|^2} +\frac{\nb(\xi)\cdot\nb'(\xi)}{|\nb(\xi)|^2} \right),
\end{aligned}
\end{equation}
and as
\begin{equation}
\mb(0)=l_{ij}\eb_{ij}, \quad \mb'(0)=\ub_j-\ub_i, \quad \nb(0)=l_{ik}\eb_{ik}, \quad \nb'(0)=\ub_k-\ub_i, \quad \widetilde{\delta\theta}_{ijk}(0)=\vartheta_{ijk},
\end{equation}
we obtain that
\begin{equation}
\begin{aligned}
\widetilde{\delta\theta}_{ijk}^\prime(0)=&-\frac{1}{\sin\vartheta_{ijk}}\left(\frac{1}{l_{ij}}\big(\eb_{ik}-(\eb_{ij}\cdot\eb_{ik})\,\eb_{ij} \big)\cdot(\ub_j-\ub_i)+\frac{1}{l_{ik}}\big(\eb_{ij}-(\eb_{ij}\cdot\eb_{ik})\,\eb_{ik} \big)\cdot(\ub_k-\ub_i)  \right)\\
&-\frac{1}{\sin\vartheta_{ijk}}\left(\frac{1}{l_{ij}}\Pb_{ij}\eb_{ik}\cdot(\ub_j-\ub_i)+\frac{1}{l_{ik}}\Pb_{ik}\eb_{ij}\cdot(\ub_k-\ub_i)\right)=-\frac{1}{l_{ij}}\,\wb_{ijk}\cdot(\ub_j-\ub_i)-\frac{1}{l_{ik}}\,\wb_{ikj}\cdot(\ub_k-\ub_i)\,;
\end{aligned}
\end{equation}
 in the last equality, we have made use of \eqref{wbb}. Then,
\begin{equation}
\widetilde{\delta\theta}_{ijk}(\xi)= \widetilde{\delta\theta}_{ijk}(0)+\widetilde{\delta\theta}_{ijk}^\prime(0)\,\xi+O(\xi^2)=\vartheta_{ijk}-\Big(\frac{1}{l_{ij}}\,\wb_{ijk}\cdot(\ub_j-\ub_i)+\frac{1}{l_{ik}}\,\wb_{ikj}\cdot(\ub_k-\ub_i)\Big)\xi+O(\xi^2),
\end{equation}
and we conclude that the linear measure of angle changes must have the form \eqref{psi}.
}

\section{Linear Theory}
\subsection{Virtual powers}
As a first step, we assume that $\Sc$ is subject to a system of external \textit{nodal forces} $\fb_i$ that are {\em dead}, that is, independent of $t$  and $\{\ub_i,\dot{\ub}_i\,|\, i\in{\mathcal N}\}$, and we define the {\em external virtual power} as follows:
\begin{equation}
\Wc^e:=\sum_{i\in\Nc}\fb_i\cdot\delta\ub_i,
\end{equation}
where $\delta\ub_i$ is the {\em virtual displacement} of the $i-$th node.

The customary next step consists in  specifying a stress system for $\mathcal S$ by duality, through the action over a chosen collection of virtual strains of a linear functional, the internal virtual power; the conclusive step, equating external and internal virtual powers subject to an appropriate quantification, yields a set of equations that are interpreted as the conditions the stress system must satisfy to balance the given load system. But, to balance it in what placement, referential or current?

 This question has an automatic answer, depending on the what collection of virtual strains is chosen.\footnote{For a discussion of this issue we refer the reader to \cite{Primer}, Section 12.} In our present case, such collection can be molded either on the collection of  exact strain rates \eqref{rcl} and \eqref{rca} or on the collection of their linearized versions \eqref{reps} and \eqref{rpsi}. The stress measures delivered by the former choice balance the given loads in the current placement of $\mathcal S$; the latter choice leads to stress measures that balance the loads in the reference placement. We here follow the second route, with a view to construct a linear theory of elastic S\&S structures. Accordingly, the form we choose for the {\em internal virtual power} is:
\begin{equation}\label{intp}
\Wc^i:=\sum_{ij\in\Ec}\sigma_{ij}\,\delta\varepsilon_{ij}+ \sum_{ijk\in\Wc}\tau_{ijk}\,\delta\psi_{ijk},
\end{equation}
where the {\em virtual strains} are:
$$
\delta\varepsilon_{ij}=\eb_{ji}\cdot(\delta\ub_i-\delta\ub_j) \quad{\rm and} \quad \delta\psi_{ijk}=\left(\frac{\wb_{ijk}}{l_{ij}}+\frac{\wb_{ikj}}{l_{ik}}\right)\cdot\delta\ub_i- \frac{\wb_{ijk}}{l_{ij}}\cdot\delta\ub_j-\frac{\wb_{ikj}}{l_{ik}}\cdot\delta\ub_k\,.
$$
 Finally, we postulate that
\begin{equation}\label{vw}
\Wc^e=\Wc^i,
\end{equation}
for all virtual displacements $\delta\ub_i$ and virtual deformations $\delta\varepsilon_{ij}$ and $\delta\psi_{ijk}$.

When in \eqref{vw} one takes $\delta\ub_i\neq\mathbf{0}$ and all the other displacements null, the \textit{equilibrium equation} of node $i$ follows:
\begin{equation}\label{eql}
\fb_i=
\sum_{\{j\in\Nc|ij\in\Ec\}} \sigma_{ij}\eb_{ji} +
\sum_{\{j,k\in\Nc|ijk\in\Wc\}}\tau_{ijk}\left(\frac{\wb_{ijk}}{l_{ij}}+ \frac{\wb_{ikj}}{l_{ik}}\right) -
\sum_{\{k,j\in\Nc|kij\in\Wc\}} \tau_{kij}\frac{\wb_{kij}}{l_{ki}}\,.
\end{equation}
In the RHS of \eqref{eql}, the first sum is over the edges stemming from node $i$; the second is over the wedges having $i$ as their head node; and the third is over the wedges having $i$ as one of their tail nodes: the  three associated resultant internal forces balance the applied external force. Satisfaction of equation \eqref{eql} for all nodes guarantees that the stress system involved balances the given loads in the given referential placement.
%
\subsection{Compatibility, equilibrium, and constitutive, operators}\label{3.2}
We now introduce a few definitions that allow for an agile manipulations of nodal displacements, strain measures and stress measures. To begin with, let $\ub$ and $\etab$ denote, respectively, the $N$-string of nodal-displacement vectors and the $(E+W)$-string of strain components:
\begin{equation}\label{compact}
[\ub]=\left[
  \begin{array}{c}
      \vdots \\
      \ub_i \\
      \vdots \\
  \end{array}
\right]\qquad\textrm{and}\qquad
[\etab]=\left[
  \begin{array}{c}
      \vdots \\
      \varepsilon_{ij} \\
      \vdots \\  \hline
      \vdots \\
      \psi_{ijk} \\
      \vdots \\
  \end{array}
\right]\,.
\end{equation}
Then, relations  \eqref{eps} and \eqref{psi2} can be written in the following compact form:
\begin{equation}\label{kincomp}
\etab=\Ab{^T}\ub,
\end{equation}
where the linear mapping $\Ab^T$ is called the \textit{kinematic compatibility operator}. Likewise, let
$\fb$ and $\chib$ denote, respectively, the $N$-string of nodal-force vectors and the $(E+W)$-string of stress components:
\begin{equation}\label{compact}
[\fb]=\left[
  \begin{array}{c}
    \vdots \\
    \fb_i \\
    \vdots \\
  \end{array}
\right]\,,\qquad
[\chib]=\left[
  \begin{array}{c}
      \vdots \\
      \sigma_{ij} \\
      \vdots \\  \hline
      \vdots \\
      \tau_{ijk} \\
      \vdots \\
  \end{array}
\right]\,.
\end{equation}
With the use of this notation,  system \eqref{eql} can be written as
\begin{equation}\label{ibrio}
\fb=\Ab \chib,
\end{equation}
where $\Ab$, the formal adjoint of the compatibility operator, is called the \textit{equilibrium operator}.

Next, let the positive {\em stiffness constants} $\kappa_{ij}$ and $\lambda_{ijk}$ characterize the linear elastic response of edge and wedge springs:
\begin{equation}\label{constitutive}
\sigma_{ij}=\kappa_{ij}\,\varepsilon_{ij}\ \ \mathrm{and} \ \ \ \tau_{ijk}=\lambda_{ijk}\,\psi_{ijk};
\end{equation}
we write this set of relations compactly:
\begin{equation}\label{constitutive2}
\chib=\Cb\etab,
\end{equation}
where $\Cb$, the {\em constitutive operator},  is a $(E+W)\times(E+W)$ diagonal matrix:
\begin{equation}\label{constitutive:diag}
[\Cb]=[\mathrm{diag}(\cdots,\kappa_{ij},\cdots | \cdots,\lambda_{ijk},\cdots)]\,.
\end{equation}
\subsection{Displacement formulation of the equilibrium problem}\label{dispfor}
We are now in a position to formulate the equilibrium problem for a typical S\&S structure in terms of displacement. This we do by  substituting into  \eqref{ibrio} the combination of equations \eqref{constitutive2} and \eqref{kincomp}; we find:
\begin{equation}\label{sti:op}
\fb=\Kb\ub\,,\quad \Kb:=\Ab\Cb\Ab^T,
\end{equation}
where $\Kb$ is the {\em linear stiffness operator}.
The corresponding displacement equilibrium equation for the $i$-th node is easily obtained by substituting \eqref{constitutive} into \eqref{eql} first, then by making use of \eqref{eps} and \eqref{psi2}:
\begin{equation}\label{sti}
\begin{split}
\fb_i=&
\sum_{\{j\in\Nc|ij\in\Ec\}} \eb_{ji}(\kappa_{ij}\eb_{ji}\cdot(\ub_i-\ub_j)) +\\
+& \sum_{\{j,k\in\Nc|ijk\in\Wc\}}\left(\frac{\wb_{ijk}}{l_{ij}}+\frac{\wb_{ikj}}{l_{ik}}\right)
\lambda_{ijk}\left(\left(\frac{\wb_{ijk}}{l_{ij}}+\frac{\wb_{ikj}}{l_{ik}}\right)\cdot\ub_i-
\frac{\wb_{ijk}}{l_{ij}}\cdot\ub_j-\frac{\wb_{ikj}}{l_{ik}}\cdot\ub_k\right) +\\
-& \sum_{\{k,j\in\Nc|kij\in\Wc\}} \frac{\wb_{kij}}{l_{ki}}
\lambda_{kij}\left(\left(\frac{\wb_{kij}}{l_{ki}}+\frac{\wb_{kji}}{l_{kj}}\right)\cdot\ub_k-
\frac{\wb_{kij}}{l_{ki}}\cdot\ub_i-\frac{\wb_{kji}}{l_{kj}}\cdot\ub_j\right)\,.
\end{split}
\end{equation}

The {\em equilibrium elastic state} $(\ub,\etab,\chib)$ of a S\&S structure is determined by looking for a solution of \eqref{sti:op} for nodal displacements, then by obtaining the strains from \eqref{kincomp} and the stresses from \eqref{constitutive2}.
In the next subsection, we shall offer a short discussion of existence and uniqueness issues.
\vskip 6pt

\begin{figure}[h]
\centering
\vskip 12pt
\includegraphics[scale=1]{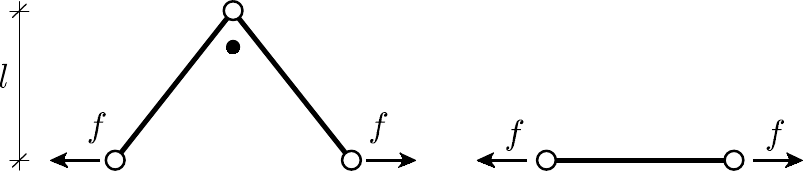}
\caption{A wedge and an edge, with spring constants $\lambda$ and $\kappa$, loaded by the same pair of forces.}
\label{pinzetta} 
\end{figure}

\noindent \textit{Remark 1.}
In a S\&S structure, energy storage in edge and wedge springs depends on the dimensionless ratio:
$$
\nu:=\frac{\kappa l^2}{\lambda},
$$
where $\kappa$ and $\lambda$ are representative spring stiffnesses and $l$ is a representative edge length.\footnote{The parameter $\nu$ was introduced in \cite{BFPG} (footnote 8) for S\&S models of carbon nanotubes.} Roughly speaking, $\nu$ measures the relative importance of edge-stretching and wedge-opening; this interpretation is made evident by considering the simple structures in Figure \ref{pinzetta},
%
where $\nu$ equals the ratio between the relative displacement of the points where forces are applied.

%
\vskip 6pt

\noindent \textit{Remark 2.} In preparation for computer calculations, it is habitual to partition the usually `big' 
stiffness matrix into `small' elementary 
blocks. We now show how this can be done for the S\&S structures we here consider.
%

%
Firstly, we single out a typical edge $ij$,  for which \eqref{sti:op} reads:
\begin{equation}\label{sti:edge:0}
\fb_{ij}=\Kb_{ij}\,\ub_{ij}\,,
\end{equation}
with
$$[\fb_{ij}]=\left[
\begin{array}{c}
\fb_i \\
\fb_j \\
\end{array}
\right]\,,\quad
[\ub_{ij}]=\left[
\begin{array}{c}
\ub_i \\
\ub_j \\
\end{array}
\right]\,,
$$
and, in view of \eqref{sti}, 
\begin{equation}\label{sti:edge}
[\Kb_{ij}]=\kappa_{ij}\left[
\begin{array}{cc}
\eb_{ij}\otimes\eb_{ij} & -\eb_{ij}\otimes\eb_{ij} \\
-\eb_{ij}\otimes\eb_{ij} & \eb_{ij}\otimes\eb_{ij} \\
\end{array}
\right]\,.
\end{equation}
Secondly, we fix attention on a typical wedge $ijk$, for which \eqref{sti:op} reads:
\begin{equation}\label{sti:wedge:0}
\fb_{ijk}=\Kb_{ijk}\,\ub_{ijk}\,,
\end{equation}
with
$$[\fb_{ijk}]=\left[
\begin{array}{c}
\fb_i \\
\fb_j \\
\fb_k \\
\end{array}
\right]\,,\quad
[\ub_{ijk}]=\left[
\begin{array}{c}
\ub_i \\
\ub_j \\
\ub_k \\
\end{array}
\right]\,,
$$
and, by inspection of \eqref{sti}, 
\begin{equation}\label{sti:wedge}
[\Kb_{ijk}]=\lambda_{ijk}\left[
\begin{array}{ccc}
(\ab+\bb)\otimes(\ab+\bb) & -(\ab+\bb)\otimes\ab & -(\ab+\bb)\otimes\bb \\
-\ab\otimes(\ab+\bb) & \ab\otimes\ab & \ab\otimes\bb \\
-\bb\otimes(\ab+\bb) & \bb\otimes\ab & \bb\otimes\bb \\
\end{array}
\right]\,,
\end{equation}
with
\begin{equation}\label{ab}
\ab:=\frac{\wb_{ijk}}{l_{ij}}\ \ \mathrm{and} \ \ \ \bb:=\frac{\wb_{ikj}}{l_{ik}}\,.
\end{equation}
Formally, we can introduce the operators $\Qb_{ij}$ and $\Qb_{ijk}$, which `extract' from $\ub$ the vectors $\ub_{ij}$ and $\ub_{ijk}$:
\begin{equation}\label{Q}
\ub_{ij}=\Qb_{ij}\,\ub\,,\qquad\ub_{ijk}=\Qb_{ijk}\,\ub\,.
\end{equation}
With these positions, we can express the external load vector $\fb$ as a sum of edge and wedge contributions:
\begin{equation}\label{f}
\fb=\sum_{ij\in\Ec}\Qb_{ij}^T\,\fb_{ij}+\sum_{ijk\in\Wc}\Qb_{ijk}^T\,\fb_{ijk}\,.
\end{equation}
Finally, by the use of \eqref{sti:edge:0}, \eqref{sti:wedge:0}, \eqref{Q} and \eqref{sti:op}, we obtain:
\begin{equation}\label{sti:op:Q}
\Kb=\sum_{ij\in\Ec}\Qb_{ij}^T\,\Kb_{ij}\,\Qb_{ij}+\sum_{ijk\in\Wc}\Qb_{ijk}^T\,\Kb_{ijk}\,\Qb_{ijk}\,.
\end{equation}
%

\subsection{Self-stresses, mechanisms, and prestresses}\label{class}
Let us recall a number of customary definitions and well-known facts about linearly elastic discrete structures in a reference placement (vid. e.g. \cite{PC86}), in a form appropriate to S\&S structures.

A {\em self-stress} is a stress vector $\chib$ which balances null loads, i.e., such that $\Ab\chib=\mathbf{0}$;  a {\em mechanism} is a displacement vector $\ub$ which induces null deformations in all edges and wedges, i.e., such that $\Ab^T\ub=\mathbf{0}$. A mechanisms is {\em nontrivial} if it is not a rigid motion of the structure (in a point space of dimension $D$, there are at most $D(D+1)/2$ independent trivial mechanisms). To focus on nontrivial mechanisms, we assume that a convenient number of external constraints forbid all rigid motions of each S\&S structure we consider.

 By definition, a self-stress vector belongs to the null space of the equilibrium operator; we denote by $S:=\textrm{dim}(\textrm{ker}(\Ab))$  the number of independent self-stresses. Moreover, since a mechanism belongs to the null space  of the compatibility operator, we take for the number of independent mechanism $M=\textrm{dim}(\textrm{ker}(\Ab^T))-D(D+1)/2$.
  Four structural types can be distinguished, according to whether $S$ and $M$ equal zero or not. These are:
%
\begin{table}[h!]
\centering
\begin{tabular}{|c|c|c|c|c|}
\hline Type & \hspace{0.19cm}1\hspace{0.19cm} & 2 &3 &4 \\
\hline  $S$ & 0 & 0 & $>0$& $>0$ \\
\hline  $M$ & 0& $>0$& 0&$>0$ \\
\hline
\end{tabular}
\caption{Self-stress\&mechanism-based structure typology.}
\label{tab1}
\end{table}

\noindent We recall that the solution of the equilibrium problem in linear elasticity exists and is unique only for structures of Type $1$ and $3$; and that, for structures of Type $2$ and $4$, a solution exists, but it is not unique, only if the load vector $\fb$ is orthogonal to all mechanisms. Counting the dimensions of the fundamental subspaces of the equilibrium and compatibility operators yields the following relation:
\begin{equation}\label{count}
3N-6-E-W=M-S
\end{equation}
 (cf. \cite{PC86}; in \eqref{count}, we have taken $D=3$).  Thus, the number of elements of $\Nc, \Ec$ and $\Wc$  determines the difference $M-S$, an important preliminary information about problem solvability:

 \noindent if $M-S=0$, then the structure at hand cannot be of Type 2 or 3; if $M-S>0$, then it cannot be of Type 1 or 3; if $M-S<0$, then it cannot be of Type 1 or 2.

A {\em prestress} is a stress vector balancing a given load vector, called {\em preload},  in a given referential placement, called {\em initial}, because it is specified before the static or dynamic response to additional loads is investigated; in view of this definition,  self-stresses are prestresses balancing null preloads.
We reinforce that the linear theory  does not always grant us a solution and that, even when it does, its predictions can be misleading, since the effect of prestress is not properly taken into account. In the next section, we shall show how a theory linearized about a prestressed initial placement -- in short, a {\em linearized theory} -- allows us to overcome the shortcomings of the linear theory, under the only assumption that the  initial placement be {\em stable}, in the sense that we make precise.

\section{Linearized Theory}

 In this section, we revert to measuring strains exactly, with the purpose of determining the \textit{tangent stiffness operator}, which specifies the linearized relation between small increments of nodal forces and small displacements from a given, possibly prestressed, equilibrium placement, whose stability we qualify.

\subsection{Tangent, elastic, and prestress, stiffness operators}\label{tso}
With a view toward giving a variational formulation of the incremental equilibrium problem, we postulate the following expression for the elastic energy stored in a S\&S structure:
\begin{equation}\label{energy}
\Uc=\frac{1}{2}\Big(\sum_{ij\in\Ec}\kappa_{ij}(\ell_{ij}-\ov \ell_{ij})^2+
\sum_{ijk\in\Wc}\lambda_{ijk}(\theta_{ijk}-\ov\theta_{ijk})^2\Big)\,,
\end{equation}
where $\ell_{ij}$ and $\ov \ell_{ij}$ are the current and rest lengths of the axial spring on edge $ij$, and where $\theta_{ijk}$ and $\ov\theta_{ijk}$ are the current and rest angles of the spiral spring on wedge $ijk$. On recalling definitions \eqref{chle} and \eqref{chan} in Section \ref{kinematics}, we see that $(\ell_{ij}-\ov \ell_{ij})$ and $(\theta_{ijk}-\ov\theta_{ijk})$ are the exact measures of, respectively, length and angle changes from a reference placement where all springs are unstrained.\footnote{Thus, as we shall see shortly, our present prescription for the stored energy has the same structure as {\em St.Venant-Kirchhoff}'s,  a widely used prescription in three-dimensional nonlinear elasticity, implying a linear dependence of the Cosserat stress from the Green-St. Venant nonlinear strain measure.}

Let $\qb$ denote the $N$-string of current position vectors of all nodes. Both the length $\ell_{ij}$ of the typical edge and the angle $\theta_{ijk}$ of the typical wedge depend on $\qb$, in a manner implicit in, respectively, $\eqref{qbb}_2$ and  \eqref{wa}.
 To lighten our notation, we rewrite \eqref{energy} as
\begin{equation}\label{energy2}
\Uc(\qb)=\frac{1}{2}\Big(\sum_{ij\in\Ec}\left(\kappa(\ell(\qb)-\ov \ell)^2\right)_{ij}+
\sum_{ijk\in\Wc}\left(\lambda(\theta(\qb)-\ov\theta)^2\right)_{ijk}\Big)\,.
\end{equation}
Consider the {\em total energy functional}:
\begin{equation}\label{tef}
\Fc(\qb):=\Uc(\qb)-\Lc(\qb), \quad \Lc(\qb):=\fb_0\cdot\ub(\qb),
\end{equation}
where $\Lc$ is the \textit{potential} of the dead load $\fb_0$. We say that $\qb_0$ is an equilibrium placement for the elastic S\&S structure at hand  if it so happens that  $\qb_0$ is a stationary point for $\mathcal F$, i.e.,
\begin{equation}\label{eqeq}
\partial_\qb \Fc(\qb_0)=\partial_\qb \Uc(\qb_0)-\fb_0=\mathbf{0}.
\end{equation}
A stationary point is a local minimum of $\mathcal F$ if it so happens that the \textit{tangent stiffness operator}
\[
\Kb_T:=\partial_\qb^2\,\Uc
\]
is {\em positive definite} at that point. Whenever this is the case, (i) we qualify the relative equilibrium placement as {\em stable}; (ii) there is a unique solution of the {\em incremental equilibrium problem}
\begin{equation}\label{incrpr}
\Kb_T\Delta\qb=\Delta\fb,
\end{equation}
where  $\Delta \qb=\qb-\qb_0$ denotes the incremental displacement induced by a small load increment $\Delta \fb=\fb-\fb_0$.



We now inspect the tangent stiffness operator carefully. The first thing we note is that it splits in two parts:

%
%
\begin{equation}\label{Ksplitt}
\Kb_T=\Kb_E+\Kb_P,
\end{equation}
where
\begin{equation}\label{Ksplit}\begin{split}
\Kb_E:=&\sum_{ij\in\Ec}\left(\kappa\,\partial_\qb \ell\otimes\partial_\qb \ell\right)_{ij}+
\sum_{ijk\in\Wc}\left(\lambda\,\partial_\qb\theta\otimes\partial_\qb\theta\right)_{ijk}\,,\\
\Kb_P:=&\sum_{ij\in\Ec}\left(\kappa(\ell-\ov \ell)\partial_\qb^2 l\right)_{ij}+
\sum_{ijk\in\Wc}\left(\lambda(\theta-\ov\theta)\partial_\qb^2\theta\right)_{ijk}\,.
\end{split}
\end{equation}
%
We call $\Kb_E$ and $\Kb_P$ the \textit{elastic stiffness operator} and the \textit{prestress stiffness operator};\footnote{In the literature, alternative nomenclatures for $\Kb_E$ and $\Kb_P$ are \textit{material stiffness operator} and \textit{geometric stiffness operator}, respectively.} their role in assessing the stability of an equilibrium placement will be briefly discussed in the next subsection.
Note that  the second of \eqref{Ksplit} can be written as
\[
\Kb_P=\sum_{ij\in\Ec}\sigma_{ij}\,\partial_\qb^2 \ell_{ij}+\sum_{ijk\in\Wc}
\tau_{ijk}\,\partial_\qb^2\theta_{ijk}\,,
\]
where
\begin{equation}\label{stresses}
\sigma_{ij}=\kappa_{ij}(\ell_{ij}-\ov \ell_{ij})\,,\qquad\tau_{ijk}=\lambda_{ijk}(\theta_{ijk}-\ov\theta_{ijk})\,,
\end{equation}
a set of {\em geometrically nonlinear} -- but {\em materially linear} -- constitutive relations (recall  footnote 6.).

It remains for us to compute the first and second derivatives with respect to $\qb$ of the strain measures. To begin with, we have from $\eqref{qbb}_2$ that
\begin{equation}\label{gradq}
\gb_{ij}:=\partial_\qb \ell_{ij}(\qb)=\frac{1}{\ell_{ij}}\Gb_{ij}\qb\,;
\end{equation}
here, $\Gb_{ij}$ is the (constant) {\em connectivity operator} of edge $ij$.
%
In a matrix form where only non-null entries are shown, relation \eqref{gradq}reads:
\begin{equation}
[\gb_{ij}]=\frac{1}{\ell_{ij}}[\Gb_{ij}\qb]=
\frac{1}{\ell_{ij}}\left[
\begin{array}{c}
\vdots\\
\qb_i-\qb_j \\
\vdots\\
\qb_j-\qb_i\\
\vdots
\end{array}
\right]=\left[
\begin{array}{c}
\vdots\\
\cb_{ji} \\
\vdots\\
\cb_{ij}\\
\vdots
\end{array}
\right]\,.
\end{equation}
Furthermore, differentiation of \eqref{gradq} yields:
\begin{equation}\label{secgradl}
\partial_\qb^2  \ell_{ij}(\qb)
=\frac{1}{\ell_{ij}}\left(\Gb_{ij}-\gb_{ij}\otimes \gb_{ij}\right).
\end{equation}
We now pass to obtain an expression for $\partial_\qb \theta_{ijk}$ and $\partial_\qb^2\theta_{ijk}$. Our point of the departure is the following consequence of \eqref{qbb} and \eqref{wa}:
\[
\cos\theta_{ijk}(\qb)=\frac{1}{ \ell_{ij}(\qb)\, \ell_{ik}(\qb)}\,(\qb_i-\qb_j)\cdot(\qb_i-\qb_k)=\frac{1}{2\, \ell_{ij}(\qb)\, \ell_{ik}(\qb)}\,\big(\ell_{ij}^2(\qb)+\ell_{ik}^2(\qb)-\ell_{jk}^2(\qb)    \big),
\]
whence we have, firstly, that
\begin{equation}\label{gradth}
\begin{aligned}
-\sin\theta_{ijk}\partial_\qb \theta_{ijk}&=\frac{1}{\ell_{ij}\ell_{ik}}\,\left(\ell_{ij}\,\gb_{ij}+\ell_{ik}\,\gb_{ik}-\ell_{jk}\,\gb_{jk}-\frac{\ell_{ij}^2+\ell_{ik}^2-\ell_{jk}^2}{2\ell_{ij}}\,\gb_{ij}-\frac{\ell_{ij}^2+\ell_{ik}^2-\ell_{jk}^2}{2\ell_{ik}}\,\gb_{ik}\right)\\
&=\frac{1}{\ell_{ij}\ell_{ik}}\Big((\ell_{ij}-\ell_{ik}\cos\theta_{ijk})\,\gb_{ij} +(\ell_{ik}-\ell_{ij}\cos\theta_{ijk})\,\gb_{ik}-\ell_{jk}\,\gb_{jk}  \Big),
\end{aligned}
\end{equation}
and, secondly, that
\begin{equation}\label{secgradt}
\begin{split}
\ell_{ij}\ell_{ik}\sin\theta_{ijk}\partial_\qb^2\theta_{ijk}=&
\,\Gb_{jk}+\Big(\frac{\ell_{ik}}{\ell_{ij}}\cos\theta_{ijk}-1\Big)\Gb_{ij}+\Big(\frac{\ell_{ij}}{\ell_{ik}}\cos\theta_{ijk}-1\Big)\Gb_{ik}\\
-&\cos\theta_{ijk}\Big(\ell_{ij}\ell_{ik}\partial_{\qb}\theta_{ijk}\otimes\partial_{\qb}\theta_{ijk}
+\frac{\ell_{ij}}{\ell_{ik}}\gb_{ij}\otimes\gb_{ij}+\frac{\ell_{ik}}{\ell_{ij}}\gb_{ik}\otimes\gb_{ik}\Big)\\
+&2\sym\big(\cos\theta_{ijk}\gb_{ij}\otimes\gb_{ik}-\sin\theta_{ijk}\,\partial_{\qb}\theta_{ijk}\otimes(\ell_{ik}\gb_{ij}+\ell_{ij}\gb_{ik})\big)
\,.\end{split}
\end{equation}
 On inserting \eqref{gradq}$-$\eqref{secgradt} in \eqref{Ksplit} and  \eqref{Ksplitt}, we obtain explicit and computable expressions for the elastic, prestress, and tangent, stiffness operators.

Interestingly, the first of these operators can be given another expression, strongly reminiscent -- and for good reasons -- of the expression derived for $\Kb$ in Remark 2, Section \ref{dispfor}. This alternative expression is arrived at by inserting \eqref{gradq} and \eqref{gradth} in  \eqref{Ksplit}$_1$. We obtain:
\begin{equation}
\Kb_E=\sum_{ij\in\Ec}\Qb_{ij}^T\,\widetilde\Kb_{ij}\,\Qb_{ij}+\sum_{ijk\in\Wc}\Qb_{ijk}^T\,\widetilde\Kb_{ijk}\,\Qb_{ijk}\,,
\end{equation}
where
\begin{equation}\label{sti:edgec}
[\widetilde\Kb_{ij}]=\kappa_{ij}\left[
\begin{array}{cc}
\cb_{ij}\otimes\cb_{ij} & -\cb_{ij}\otimes\cb_{ij} \\
-\cb_{ij}\otimes\cb_{ij} & \cb_{ij}\otimes\cb_{ij} \\
\end{array}
\right]\,,
\end{equation}
and
\begin{equation}
[\widetilde\Kb_{ijk}]=\lambda_{ijk}\left[
\begin{array}{ccc}
(\widetilde\ab+\widetilde\bb)\otimes(\widetilde\ab+\widetilde\bb) & -(\widetilde\ab+\widetilde\bb)\otimes\widetilde\ab & -(\widetilde\ab+\widetilde\bb)\otimes\widetilde\bb \\
-\widetilde\ab\otimes(\widetilde\ab+\widetilde\bb) & \widetilde\ab\otimes\widetilde\ab & \widetilde\ab\otimes\widetilde\bb \\
-\widetilde\bb\otimes(\widetilde\ab+\widetilde\bb) & \widetilde\bb\otimes\widetilde\ab & \widetilde\bb\otimes\widetilde\bb \\
\end{array}
\right]\,,\quad \widetilde\ab:=\frac{\widetilde\wb_{ijk}}{\ell_{ij}}\ \ \mathrm{and} \ \ \ \widetilde\bb:=\frac{\widetilde\wb_{ikj}}{\ell_{ik}}\,.
\end{equation}
We see that, if the current and referential placements are identified, as is done within all {\em linear} theories of elasticity, then $\Kb_P$ vanishes and $\Kb_E$ (a stiffness operator defined within a {\em linearized} elasticity theory) reduces to $\Kb$. Just as $\Kb$, $\Kb_E$ is {\em positive semi-definite}, the associated quadratic form vanishing `along mechanisms' only ($\Kb_E\ub=\mathbf{0}$ for all $\ub$ such that $\widetilde{\Ab}^T\ub=\mathbf{0}$).

\subsection{Prestress stability and superstability}\label{presss}
%


Within the present linearized theory, we have already qualified the stability of an equilibrium placement $\qb_0$  in terms of the positivity of the tangent stiffness operator at that placement. Here we investigate the signature of both $\Kb_E(\qb_0)$ and $\Kb_P(\qb_0)$.

Firstly, we consider the cases when the system at hand admits in placement $\qb_0$ some mechanisms (that is to say, there are some displacements $\ub$ for which $\Kb_E(\qb_0)\ub=\mathbf{0}$). In such cases, it may happen that the prestress balancing the preload $\fb_0$ in placement $\qb_0$ guarantees   positivity of $\Kb_T(\qb_0)$ (note that, for $\ub$ a mechanism, $\Kb_P(\qb_0)\ub=\Kb_T(\qb_0)\ub$). Accordingly, we say that a S\&S structure is {\em prestress stable} in a given equilibrium placement if the prestress stiffness operator evaluated at that placement is positive along all mechanisms the structure admits in that placement, that is, if
\begin{equation}
\Kb_P\ub\cdot\ub>0\,,\quad \textrm{for all}\;\ub\;\textrm{such that}\;\widetilde\Ab^T\ub=\mathbf{0}
\end{equation}
(cf. \cite{M13,C99,P90,CP91}).
Secondly, we say that a S\&S elastic structure is {\em superstable} at an equilibrium placement $\qb_0$ if
$\Kb_P(\qb_0)$ is positive along mechanisms and nonnegative otherwise.

\vskip 6pt

\noindent \textit{Remark 3.} If a S\&S elastic structure sits in a prestress stable, but not superstable, equilibrium placement, then the prestress stiffness operator might turn out to be negative along some displacements  (cf. \cite{C99,ZO07}). It is also important to realize that prestress can noticeably affect a structure's response even from a stable placement where no mechanisms are possible.

\vskip 6pt

\noindent \textit{Remark 4.}
The self-stress\&mechanism-based classification criterion proposed in Section \ref{class} is applicable also to structures in a prestressed placement, as we here exemplify by looking at two isomers of cyclohexane (C$_6$H$_{12}$), a most studied molecule whose energy landscape is well understood.

\begin{figure}[h]
\centering
\includegraphics[scale=.75]{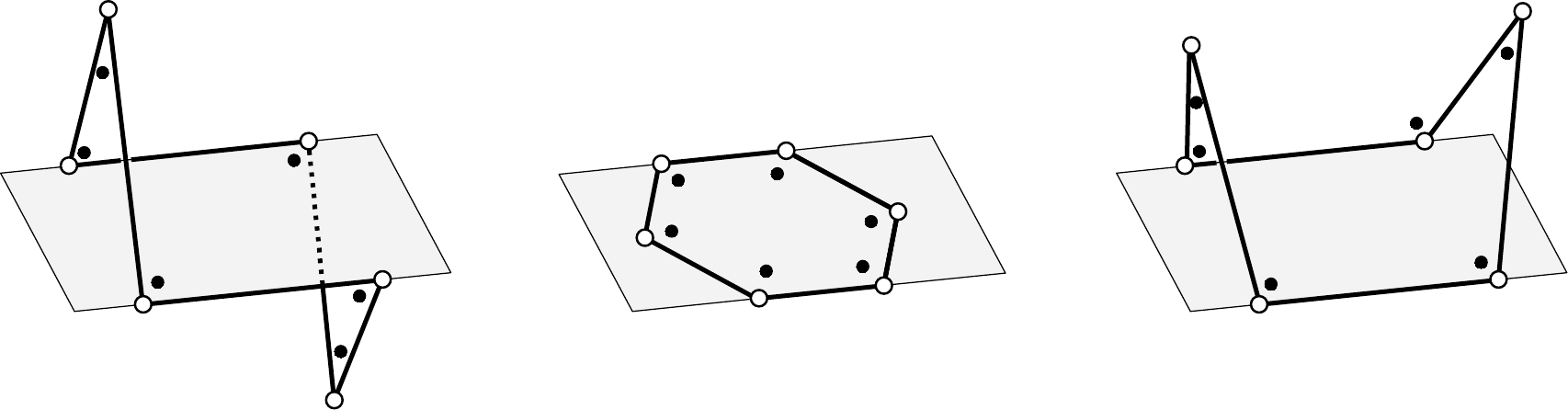}
\caption{The cyclohexane ring, in the flat (center), chair (left), and boat (right) placements.}
\label{hexagon} 
\end{figure}

From a purely mechanical viewpoint, these isomers -- the `chair' and the `boat'  (Figure \ref{hexagon}) --
can be seen as two equilibrium placements of one and the same S\&S elastic structure, whose flat placement (were it observed) would look like a  hexagonal ring with the C atoms at the corners, each C atom having two H atoms attached to it.

\begin{figure}[h]
\centering
\includegraphics[scale=1]{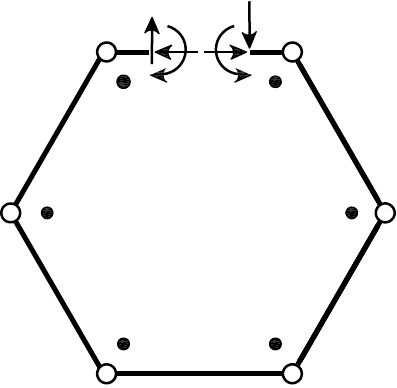}
\caption{Internal actions inducing self-stress states in a flat cyclohexane ring.}
\label{hex_ss} 
\end{figure}

The chair placement is known to be a global energy minimizer, the boat a local maximizer.\footnote{There are also two other isomers, the `half-chair' and the `twist-boat'; the former maximizes the energy, the latter realizes a local minimum \cite{NB}.} An inspection of  the null spaces of the equilibrium and compatibility operators for the chair placement reveals that they are both trivial ($S=M=0$): according to Table \ref{tab1}, the chair is a Type 1 structure.   As to the unphysical flat placement, such an inspection shows that $S=M=3$: in fact, as suggested in Figure \ref{hex_ss},
three independent self-equilibrated systems of internal actions are possible; moreover, the three possible independent mechanisms are visualized by thinking of situations when three of the nodes are kept fixed whereas the other three are made move out of the plane. Thus, according to Table \ref{tab1}, the type of a flat cyclohexane ring is 4. The same is true for the boat isomer, for which $S=M=1$. However, a flat  placement can be prestress stable (for example, when  edge springs are not stressed and wedge rest angles are all equal and larger than $2\pi/3$), due to the positive contribution of the prestress stiffness operator, whereas the boat is not stable, since the contribution of that operator is always null.\footnote{Our classification of the chair and boat placements is consistent with that given by Whiteley \cite{Whit}, who adopted a purely kinematical approach; in Whiteley's language, the former placement is `rigid', the latter`flexible'.}


%
\subsection{Motion equations}
On introducing the kinetic energy:
\begin{equation}
\Kc(\qb,\dot\qb)=\frac{1}{2}\,\dot{\qb}\cdot\Mb(\qb)\dot{\qb},
\end{equation}
where $\Mb$ is the \textit{mass operator}, the nonlinear motion equations are given by
$$
\Mb\ddot\qb+\widetilde\Ab(\qb)\chib(\qb)=\mathbf{0},
$$
where  $\widetilde\Ab (\qb)$ is the \textit{equilibrium operator} in the current placement; their linearized version about an equilibrium placement $\qb_0$  has the form:
\begin{equation}
\Mb(\qb_0) \ddot{\ub}+\Kb_T(\qb_0)\ub=\mathbf{0};
\end{equation}
note the role of the tangent stiffness operator.
In the simplest case when structural masses are lumped in the nodes,  $\Mb$ takes the form of an $N\times N$ block diagonal matrix, whose $i$-th block is  $m_i \Ib_{D}$,  the product of the mass $m_i$ of the $i$-th node and  the $D$-dimensional identity $\Ib_{D}$.


\section{Examples}
This section is devoted to illustrate by two examples what our S\&S structure theory can and cannot do. The first example makes patent the effect of prestress on the load-displacement relationship, an effect that no linear theory can capture. The second is numerical: by the use of an {\em ad hoc} code implemented by one of us \cite{M13b}, we compute the static response to loads of a prestressed S\&S structure of some complication within the linear, linearized, and geometrically nonlinear theories we propose.

We do not specify units for the mechanical and geometrical quantities we manipulate, because our purposes are purely qualitative.  In both examples, the static analyses in the large-displacement regime  are made by a standard iterative Newton method.
We sometime refer to solutions obtained in this way as {\em exact}, even if this is true only in a numerical sense, that is, up to numerical approximations and round-up errors.

\subsection{A tripod}
\begin{figure}[h]
\centering
\includegraphics[scale=1]{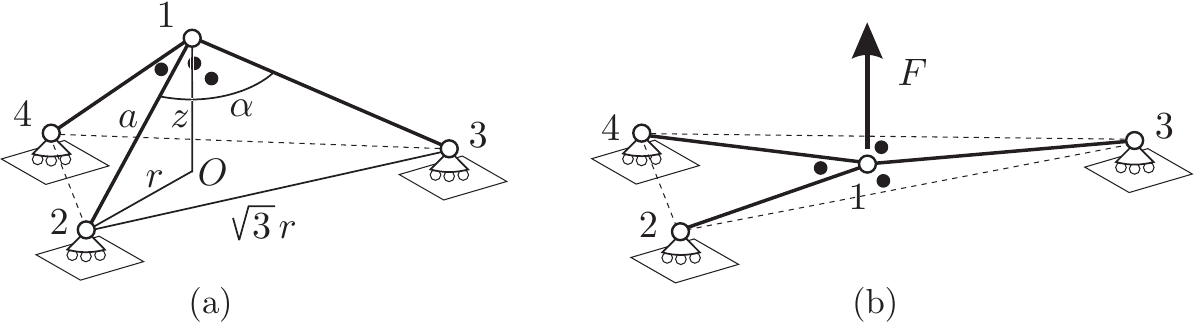}
\caption{A tripod, composed by three edges, $12$, $13$, $14$, and three wedges, $123$, $134$, $124$.}
\label{tripod01} 
\end{figure}

Let us consider an S\&S tripod with three edges $12$, $13$, and $14$, and three wedges, $123$, $134$,  and $124$ (Fig.\,\ref{tripod01}(a)). In a generic placement, nodes $2$, $3$ and $4$ are located on a horizontal plane, at the vertices of an equilateral triangle, and node $1$ is located on the vertical axis passing through the center $O$ of the triangle; nodes $2$, $3$ and $4$ can slide freely on the horizontal plane.
We are interested in computing the load-displacement relationship when an initially flat tripod is loaded by  vertical force $F$ applied at node $1$ (Fig.\,\ref{tripod01}(b)). We assume that the tripod's mechanical response is completely symmetric with respect to the vertical axis through $O$, and we choose $r$ and $z$ as geometrical parameters (Fig.\,\ref{tripod01}(a)). Then, all edges have length $a$, with
\begin{equation}\label{asquare}
a^2=r^2+z^2\,,
\end{equation}
%
and all wedges have angle $\alpha$, with
%
%
\begin{equation}\label{alphafun}
\alpha=\arccos\left(1-\frac{3r^2}{2a^2}\right)\,.
\end{equation}
%
%

The total energy functional \eqref{tef} takes the following form:
\begin{equation}\label{en:tripod}
\Uc(r,z)-Fz=\frac{1}{2}\big(3k(a(r,z)-\ov{a})^2+3\lambda(\alpha(r,z)-\ov{\alpha})^2\big)-Fz\,,
\end{equation}
where $k$ and $\lambda$ denote the stiffnesses of edge and wedge springs, whose rest length and angle are $\ov{a}$ and $\ov{\alpha}$. The equilibrium equations \eqref{eqeq} become:
\begin{equation}\label{den}
\left\{
\begin{array}{c}
3k(a-\ov{a})\,a,_r+3\lambda(\alpha-\ov{\alpha})\,\alpha,_r=0\,,
 \\[10pt]
3k(a-\ov{a})\,a,_z+3\lambda(\alpha-\ov{\alpha})\,\alpha,_z-F=0\,,
\end{array}
\right.
\end{equation}
which can be rewritten as follows,
\begin{equation}\label{eql:tripod:stress}
\left\{
\begin{array}{c}
\displaystyle \sigma\,\frac{r}{a}+\tau\,\frac{r(1+2\cos\alpha)}{a^2\sin\alpha}=0\,,
\\[10pt]
\displaystyle \sigma\,\frac{z}{a}-2\tau\,\frac{z(1-\cos\alpha)}{a^2\sin\alpha}=F\,.
\end{array}
\right.
\end{equation}
%
%
where
\begin{equation}\label{costine}
\sigma=k(a-\ov{a}),\quad \tau=\lambda(\alpha-\ov{\alpha}).
\end{equation}
We see that, for the tripod to be in equilibrium in the flat placement, $F$ must be null (because of $\eqref{eql:tripod:stress}_2$) and $\alpha=2\pi/3$ (because of \eqref{asquare} and \eqref{alphafun}); consequently, due to  $\eqref{eql:tripod:stress}_1$ and $\eqref{costine}_1$, $a=\ov{a}$. Then, initially, edges have length $r_0$ and wedges angle $\alpha_0=2\pi/3$, so that $z_0=0$; wedges  are prestressed if $\ov{\alpha}\neq 2\pi/3$, and we set: \[\tau_0=\lambda(2\pi/3-\ov{\alpha}).\] Moreover, in preparation to solving system \eqref{eql:tripod:stress}-\eqref{costine} numerically for $F=F(z)$, we compute the tangent stiffness operator (that is, the Hessian of $\mathcal U$) at $(r_0,0)$. We find, preliminarly,
\begin{equation}\label{d:at0}
\begin{split}
\left. a,_r\right|_{(r_0,0)}=1\,,\quad\left. a,_z\right|_{(r_0,0)}=0 \,, \quad
\left.\alpha,_r\right|_{(r_0,0)}=\left.\alpha,_z\right|_{(r_0,0)}=0\,,
\\
\left. a,_{rr}\right|_{(r_0,0)}=\left. a,_{rz}\right|_{(r_0,0)}=0 \,, \quad
\left.\alpha,_{zz}\right|_{(r_0,0)}=-2\sqrt{3}\,r_0^{-2}\,;
\end{split}
\end{equation}
with this, we conclude that
\begin{equation}\label{dden:at0}
\Kb_T(r_0,0)=\left[\begin{array}{cc}{\mathcal U},_{rr}(r_0,0) & 0 \\ 0 & {\mathcal U},_{zz}(r_0,0)\end{array}\right]=
\left[\begin{array}{cc}3k& 0 \\ 0 & -6\sqrt 3\,\tau_0/r_0^{2}\end{array}\right]\,.
\end{equation}
%
%
Thus, the tangent stiffness operator is positive definite if and only if $\,\ov{\alpha}>2\pi/3$, that is, wedges are prestressed with $\tau_0<0$.

%
%

Figure \ref{load:disp:tripod} shows two plots of the load-displacement relationship: 
\begin{figure}[h]
\centering
\includegraphics[scale=.81]{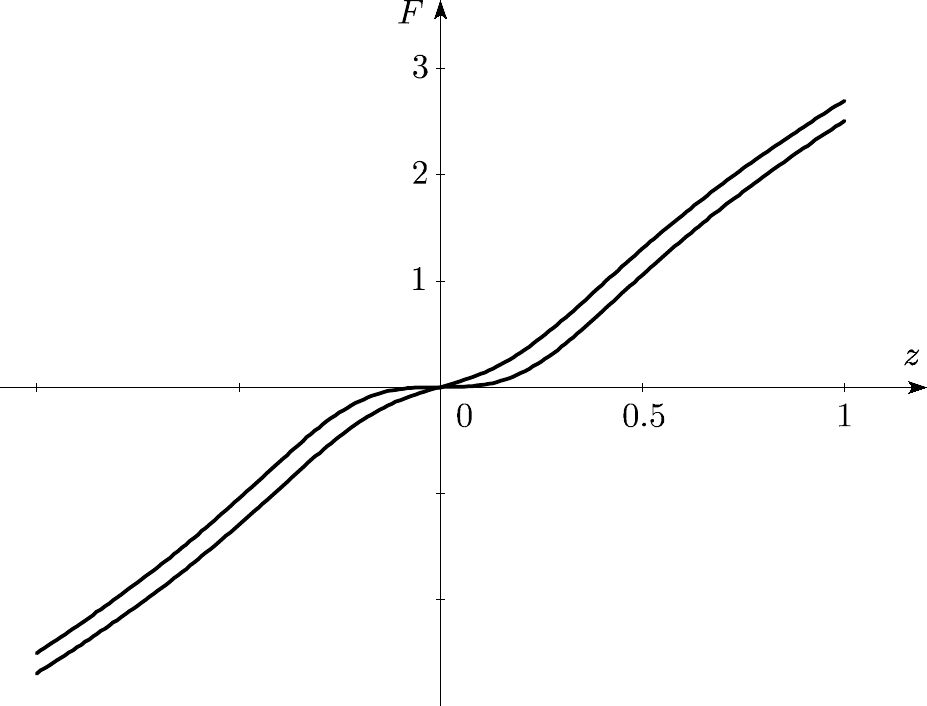}
\caption{Load-displacement curves for the flat tripod. The slope at the origin is null if the system is not prestressed, positive if the wedge springs are compressed initially.}
\label{load:disp:tripod} 
\end{figure}
the one with null slope at the origin obtains when  $\ov{\alpha}=2\pi/3$; for the other, whose slope at the origin is positive, $\ov{\alpha}=1.05(2\pi/3)$. Needless to say, the slope at the origin measures the initial stiffness, and is equal to $\Uc,_{zz}(r_0,0)$, as given in \eqref{dden:at0}. We see that, if wedges are conveniently prestressed, a flat tripod has a nonvanishing initial bending stiffness, even if $\alpha,_r(r_0,0)$ is null. The solid plots are drawn  point-by-point,  picking a value for $z$ in \eqref{eql:tripod:stress}$_1$, then solving that equation for $r$ numerically, then computing $F$ from \eqref{eql:tripod:stress}$_2$.  The following are the chosen parameter values: $r_0=\ov{a}=1$, $\kappa=1$, $\lambda=1$. 

%

The effects of prestress are evident in dynamics as well.
Suppose that all nodes have the same mass $m$, and take $F=0$. 
 The tripod's kinetic energy is:
\begin{equation}
\Kc=\frac{1}{2}m\big(3\dot{r}^2+\dot{z}^2 \big);
\end{equation}
the mass operator is:
\begin{equation}
[\Mb]=\left[
\begin{array}{ccc}
3m & 0 \\ [0.2cm]
0 & m
\end{array}
\right]\,;
\end{equation}
 the equation of motion are:
\begin{equation}\label{dintrip}
\left\{
\begin{array}{c} m\ddot{r}+\kappa(a-\bar{a})a,_r+\lambda(\alpha-\bar{\alpha})\alpha,_r=0,\\ [0.2pt]
m\ddot{z}+3\kappa(a-\bar{a})a,_z+3\lambda(\alpha-\bar{\alpha})\alpha,_z=0,\\
\end{array}
\right.
\end{equation}
and have the following linearized version:
%
\begin{equation}\label{lin:dyn:tripod}
\left\{
\begin{array}{c}
 m\ddot{r}+\kappa\, r=0,\\
 m\ddot{z}+ 6\sqrt 3\,\tau_0/r_0^{2}\,z=0.
 \end{array}
 \right.
\end{equation}
Note that \eqref{lin:dyn:tripod} is physically meaningful meaningful if the tripod is prestressed, with $\tau_0=\lambda(2\pi/3 -\ov\alpha)<0$; if this is the case, it behaves as a pair of independent harmonic oscillators, with angular frequencies:
\begin{equation}
\omega_1^2=\frac{\kappa}{m}, \quad \omega_2^2=6\sqrt{3}\,\lambda\,\frac{|2\pi/3 -\ov\alpha|}{mr_0^2}\,.
\end{equation}
We see that $\omega_1$ depends on the edge stiffness parameter $\kappa$, while $\omega_2$ depends not only on the wedge stiffness parameter $\lambda$ but also on the geometric stiffness parameter  $|2\pi/3 -\ov\alpha|/r_0^2$. We see that the frequency ratio $\omega_1/\omega_2$ depends on the dimensionless parameter $\nu$ introduced in Remark 1, Section \ref{3.2}; in this example, $\nu=\kappa r_0^2/\lambda$.
%

%
\subsection{A warehouse}\label{ex3}
\begin{figure}[h!]
\centering
\includegraphics[scale=0.8]{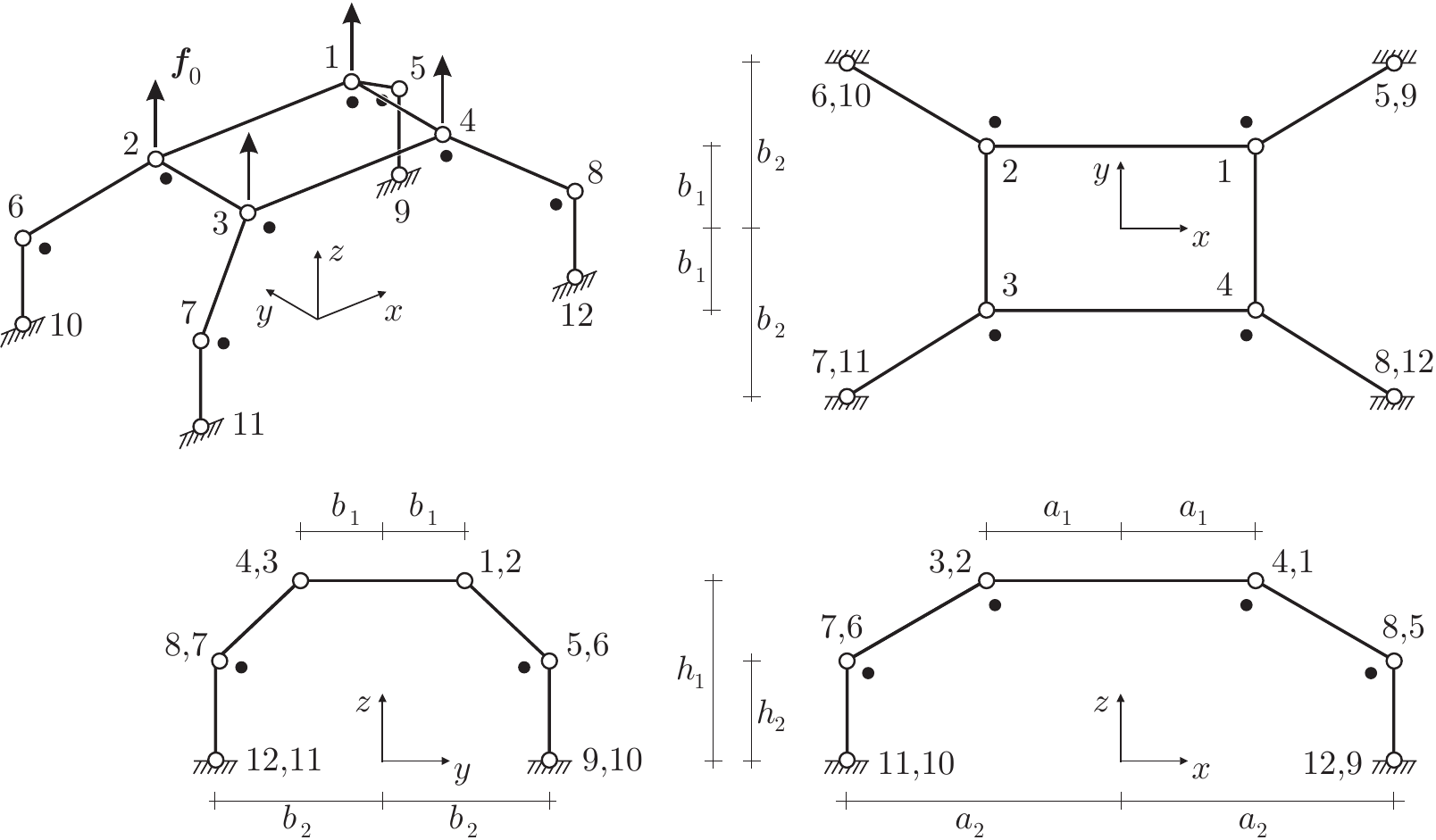}
\caption{The initial placement of the warehouse analyzed in Section \ref{ex3}; note the preload system $\fb_0$.}
\label{example3} 
\end{figure}

Consider the S\&S structure shown  in Fig.\,\ref{example3} -- a sort of warehouse -- with 12 nodes, 12 edges, and 8 wedges, all listed in  Table \ref{ex3table1}. Nodes $9, 10, 11$ and $12$ are fixed on the same horizontal plane $z=0$; the geometrical dimension are:
$$a_1=200\,,\quad a_2=400\,,\quad b_1=100\,,\quad b_2=200\,,\quad h_1=100\,,\quad h_2=200\,.$$
All edge springs have equal stiffness $\kappa$, and all wedge springs have equal stiffness $\lambda$, with
$$\kappa=1\,,\quad \lambda=\kappa b_1^2=10^4\,.$$
\noindent
In its initial placement, this is a Type $2$  structure, with $S=0$ and $M=4$, preloaded by four equal vertical forces of magnitude $F=50$, acting on nodes $1$, $2$, $3$, $4$; the preload, that we denote by $\fb_0$, is balanced by the prestress $\chib_0$ (see Table \ref{ex3table1}), and is orthogonal to all mechanisms.

\begin{figure}[h!]
\centering
\includegraphics[scale=0.8]{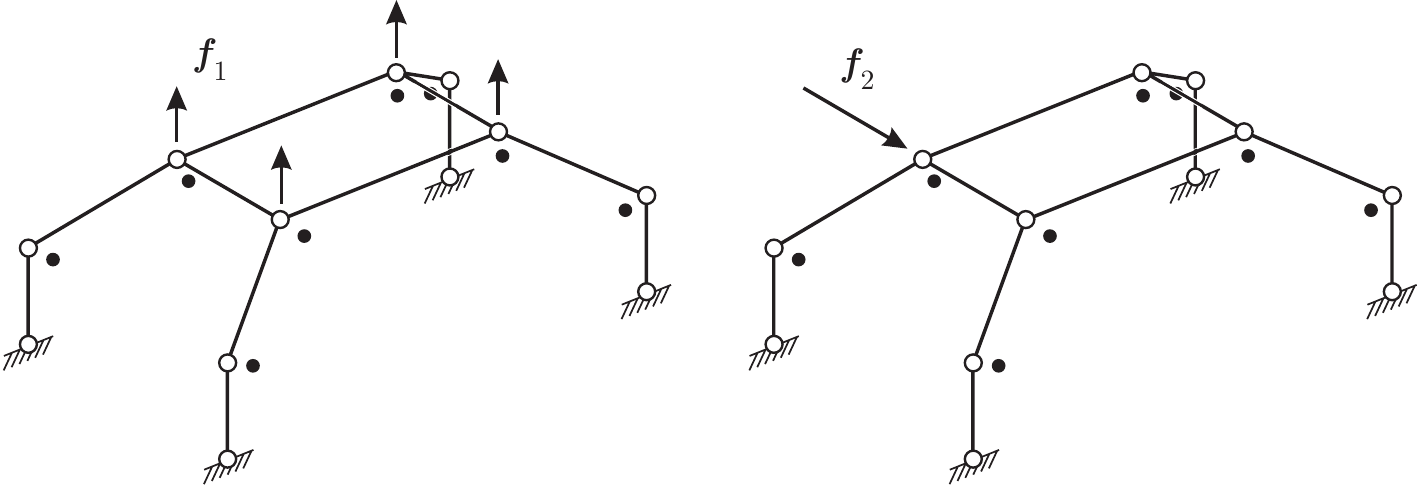}
\caption{The additional load systems $\fb_1$ and $\fb_2$.}
\label{ex3loads} 
\end{figure}

We consider the two systems of additional loads shown in Fig.\,\ref{ex3loads}, namely, an incremental system  $\fb_1=0.1\,\fb_0$ and a single horizontal force $\fb_2$ in the direction of the $y$-axis, applied at node $2$.
%
\noindent Since the first additional load, just as the preload, is orthogonal to all mechanisms, the system $\Kb_E\ub=\fb_1$ can still be solved for $\ub$; in Table \ref{ex3table2}, the only solution which is orthogonal to all mechanisms is recorded. By solving the linearized problem \eqref{incrpr} for $\Delta\qb$  with $\Delta\fb=\fb_1$  (Table \ref{ex3table2}, `linearized' column), we see that the prestress contribution $\Kb_P$ to $\Kb_T$ is so relevant that  the linear theory gives far-off results;  stresses are then computed in terms of linearized strains. Finally, a numerically exact large-displacement solution has been obtained, with the use of a FE approach. The results of all computations are collected in Tables \ref{ex3table1} and \ref{ex3table2}.

In the case of the second additional load, a linear setup precludes existence of a solution. Both the solution of the linearized equations and the solution in the large-displacement regime are recorded in  Tables \ref{ex3table1} and \ref{ex3table2}. We see that the two solutions are comparable, although their difference is not negligible. This was also true in the first load case.
\begin{table}[h!]
\centering
\includegraphics[scale=0.9]{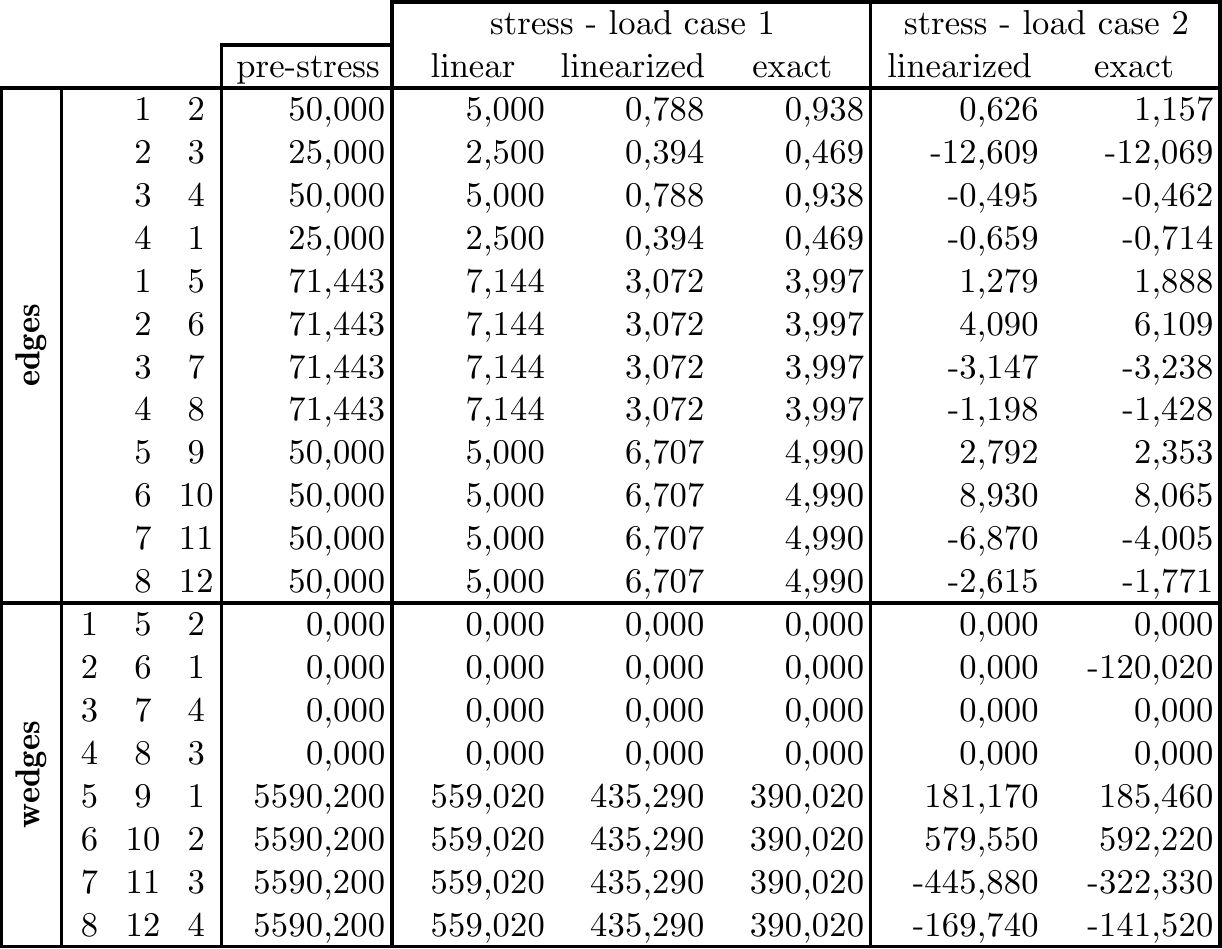}
\caption{Prestresses and stresses for the two load cases.}
\label{ex3table1} 
\end{table}
\begin{table}[h]
\centering
\includegraphics[scale=0.9]{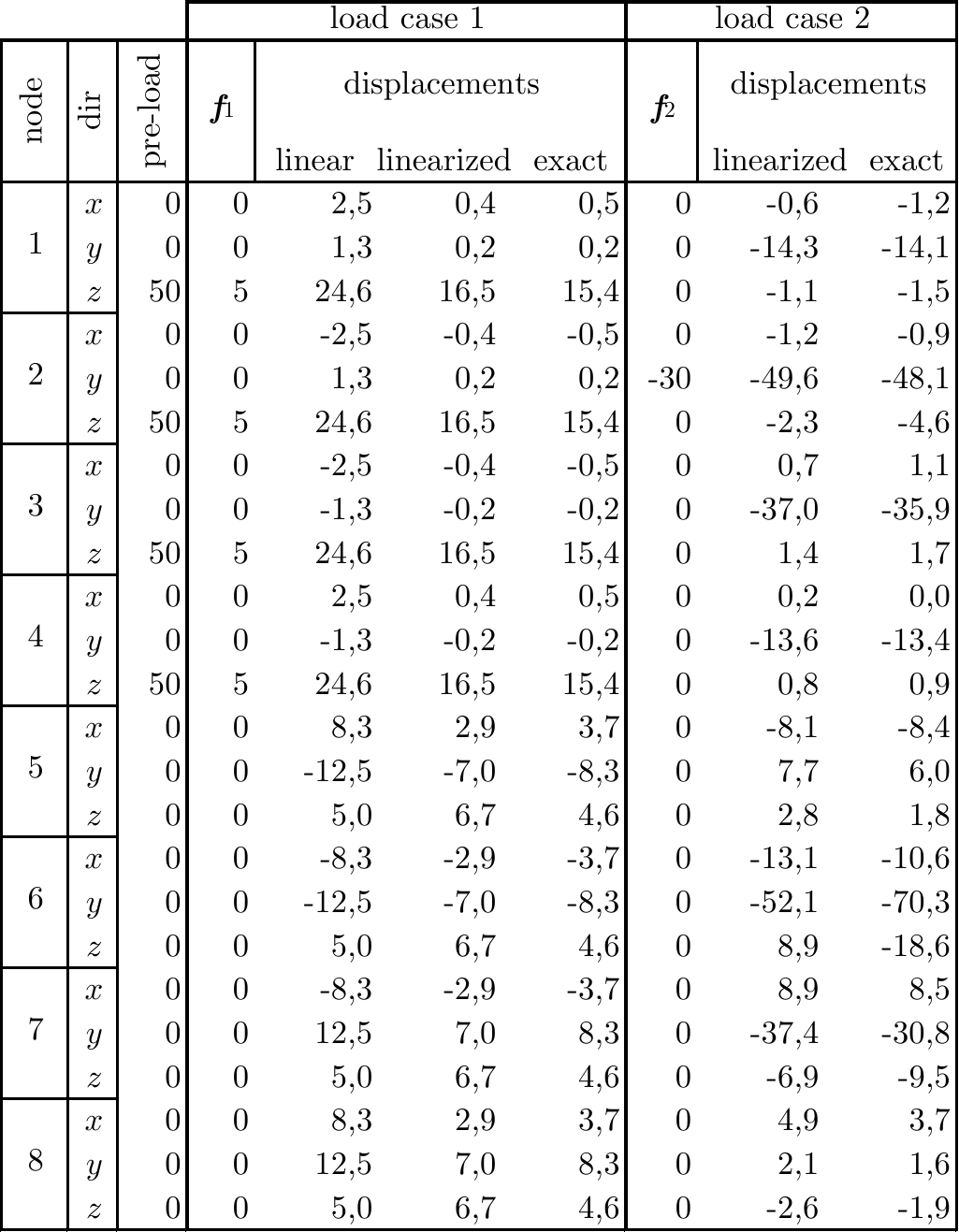}
\caption{Node displacements for the two load cases.}
\label{ex3table2} 
\end{table}
\vspace{1cm}

\section{Conclusions and directions of future work}
A  theory of stick-and-spring elastic structures  accounting for prestress has been presented, with potential application to graphene, nanotubes and other nanostructures, such as viral capsids and protein complexes.  Within such a theory, elastic energy is stored because of changes in edge lengths and wedge angles, expressed in terms of nodal displacements. A pair of exact \textit{non-linear} strain measures has been introduced, as well as their \textit{linearized} versions. The theory has been presented in three setups, referred to, respectively, as linear, linearized, and nonlinear; both natural and prestressed referential and initial placements can be accommodated, and their stability qualified. 
%
%
%
The model has been implemented in a numerical code, that can handle any of the three theoretical setups, and has been used, in particular, to provide explicit solutions of the structural problems considered in the concluding examples.

Both the model and its numerical implementations lend to various generalizations; we are especially interested in pursuing the following ones:
\begin{enumerate}[(i)]
\item \textit{Enriched kinematics}: changes in dihedral and solid angles can be taken into account, both exactly and through definition of additional linearized strain measures. In order to do that, two different types of \textit{hyperedges}, each of which individuated by four nodes, are to be considered: in the case of a {\em dihedral hyperedge}, two of the nodes serve the purpose of defining an edge, while the other two define the planes of the relative dihedral angle; in the case of a {\em solid hyperedge}, one head node and three tail nodes identify a solid angle. Then, it would not be difficult to define in terms of nodal coordinates appropriate exact and linearized hyperedge strain measures, and develop the new, kinematically enriched theory along the lines of the old one.

\item
\textit{Enriched kinetic energy}: structural masses could be lumped not only at the  nodes, but also along edges and wedges; how to assign masses, concentrated or diffused, to the various structural elements could be done in a application-specific manner, that is, in a manner that takes into account the nature of the physical nanostructures the model aims to mimic.
\end{enumerate}

We are currently developing these lines of research with a view to
a better understanding of the statical and dynamical behavior of nanostructures.

\section*{Acknowledgements}
The work of AF and AM was part of their activities within the framework of the 2013 Project ``Modelli di Nanostrutture e Biomateriali in Meccanica Molecolare'' of INdAM--GNFM.


\begin{thebibliography}{00}
	
\bibitem{Ab} G.C. Abell, Empirical chemical pseudopotential theory of molecular and metallic bonding, Phys. Rev. B 31, 6184-6196 (1985) 

\bibitem{BFPG} C. Bajaj, A. Favata, P. Podio-Guidugli,
On a nanoscopically-informed shell theory of single-wall carbon nanotubes, Europ. J. Mech. A/Solids,  42 (2013) 137-157

\bibitem{Br} D.W. Brenner, Empirical potential for hydrocarbons for use in simulating the chemical vapor deposition of diamond films, Phys. Rev. B 42, 9458-9471 (1990) 

\bibitem{Br2} D.W. Brenner, O.A. Shenderova, J.A. Harrison, S.J. Stuart, B. Ni, S.B. Sinnott, A second-generation reactive empirical bond order (REBO) potential energy expression for
hydrocarbons, J. Phys.: Condens. Matter, 14 (2002) 783-802 

\bibitem{CP91} C.R. Calladine, S. Pellegrino, First-order infinitesimal mechanisms, Int. J. Solids Struct. 27 (1991) 505-515
	


 \bibitem{Cha1} T. Chang,  {A molecular based anisotropic shell model for single-walled
 carbon nanotubes.} J. Mech. Pys. Sol. 58, 9, 1422-1433 (2010)

\bibitem{CGG1} T. Chang, J. Geng, X.  Guo, Chirality- and size- dependent elastic properties of singlewalled
carbon nanotubes. Appl. Phys. Lett. 87 (2005), 251 929


\bibitem{CGG} T. Chang, J. Geng, X. Guo, Prediction of chirality- and size-dependent elastic properties of single-walled carbon nanotubes via a molecular mechanics model, Proc. R. Soc. A 2006 462, 2523-2540


\bibitem{CG}T. Chang, H. Gao, Size-dependent elastic properties of a single-walled carbon nanotube via a molecular mechanics model, J. Mech. Phys. Solids, 51 (2003), 1059-1974

\bibitem{C99}  R. Connelly, Tensegrity structures: Why are they stable? in: \textit{Rigidity theory and applications} (Thorpe and Duxbury eds.), Kluwer 1999, pp. 47-54

\bibitem{FPG}  A. Favata, P. Podio-Guidugli, A shell theory for chiral carbon nanotubes, http://arxiv.org/abs/1306.6828, submitted (2013)

\bibitem{Fin} M.W. Finnis, J.E. Sinclair, A simple empirical N-body potential for transition metals, Phil. Mag. A 50(1): 45 (1984)

\bibitem{GC} J. Geng, T. Chang, {Nonlinear stick-and-spring model for predicting mechanical behavior of single-walled carbon
nanotubes}, Phys. Rev. B 74:245428 (2006)

\bibitem{Gerg1} S.K. Georgantzinos, G.I. Giannopoulos,
N.K. Anifantis, Numerical investigation of elastic mechanical properties of graphene structures, Materials \& Design,31, 10 (2010),  4646-4654

\bibitem{Gerg} S.K. Georgantzinos, G.I. Giannopoulos, D.E. Katsareas,
P.A. Kakavas, N.K. Anifantis, Size-dependent non-linear mechanical properties of graphene nanoribbons, Comput. Mat. Sci., 50, 7 (2011), 2057-2062


\bibitem{Gian}  G.I. Giannopoulos, I.A. Liosatos, A.K. Moukanidis, Parametric study of elastic mechanical properties of graphene nanoribbons by a new structural mechanics approach, Physica E, 44 (2011), 1, 124-134

\bibitem{Gian1}  G.I. Giannopoulos, Elastic buckling and flexural rigidity of graphene nanoribbons by using a unique translational spring element per interatomic interaction, Comput. Mat. Sci., 53, 1 (2012), 388-395



\bibitem{Meo} M. Meo, M. Rossi, Prediction of Young's modulus of single wall carbon nanotubes by molecular-mechanics based finite element modelling, Compos. Sci. Technol. 66 (2006) 1597-1605

\bibitem{Merli} R. Merli, C. L\'azaro, S. Monl\'eon, A. Domingo, A molecular structural mechanics model applied to the static behavior of single-walled
Carbon nanotubes: New general formulation. Computers \& Structures (in Press. http://dx.doi.org/10.1016/j.compstruc.2012.11.023)

\bibitem{M13} A. Micheletti, Bistable regimes in an elastic tensegrity system, Proc. R.
Soc. A, 27 (2013), 2154, 20130052

\bibitem{M13b} A. Micheletti, A computational code for large-displacement static and dynamic analyses of prestressed elastic stick\&spring structures, Technical report, Department of Civil Engineering and Information Engineering, University of Rome TorVergata (2013).

\bibitem{NB} D.J. Nelson, C.N. Brammer, Toward consistent terminology for cyclohexane conformers in introductory organic chemistry, J. Chem. Ed., 88(3), 293-294 (2011)

\bibitem{PC86} S. Pellegrino, C.R. Calladine,  Matrix analysis of statically and kinematically indeterminate frameworks Int. J. Solids Struct. 22 (1986) 409-428

\bibitem{P90} S. Pellegrino, Analysis of prestressed mechanisms, Int. J. Solids Struct. 26 (1990) 1329-1350

\bibitem{Primer}P. Podio-Guidugli, { A Primer in Elasticity},
Kluwer (2000)

\bibitem{Ba} A.J. Radez, C. Chennubothla, L.-W. Yang, I. Bahar, The Gaussian Network Model: theory and applications. In: Cui Q, Bahar I, editors. \textit{Normal Mode Analysis. Theory and Applications to Biological and Chemical Systems}. Taylor \& Francis Group, London: Chapman \& Hall/CRC; 2006. p. 41-64

\bibitem{SWG} S.A. Edwards, J. Wagner, F. Gr\"ater, Dynamic prestress in a globular protein, PLoS Comput Biol., 8(5): e1002509 (2012)

\bibitem{SL} L. Shen, J. Li, {Transversely isotropic properties of single-walled carbon nanotube}. Phys. Rev. B 69:045414 (2004)

\bibitem{SPPG}
X. Shi, B. Peng, N.M. Pugno, H. Gao, Stretch-induced softening of bending rigidity in graphene, Appl. Phys. Lett. 100, 191913 (2012)

\bibitem{Ter} J. Tersoff, New empirical approach for the structure and energy of covalent systems, Phys. Rev. B., 6991-7000 (1988) 

\bibitem{Wang} Q.Wang, Effective in-plane stiffness and bending rigidity of armchair and zigzag carbon nanotubes. Int. J. Solids Struct. 41 (2004) 5451-5461

\bibitem{Whit} W. Whiteley, Rigidity of molecular structures: geometric and generic analysis, in: \textit{Rigidity theory and applications} (Thorpe and Duxbury eds.), Kluwer 1999, pp. 21-46


\bibitem{Xiao} J. Xiao, B. Gama, J. Gillespie Jr., An analytical molecular structural mechanics model for the mechanical properties of carbon nanotubes, Int. J. Solids Struct. 42 (2005) 3075-3092

\bibitem{ZO07} J.Y. Zhang, M. Ohsaki, Stability conditions for tensegrity structures, Int. J. Solids Struct. 44 (2007) 3875-3886

\end{thebibliography}
\end{document}